\documentstyle[12pt]{article}
\def\ZZ{{\mathchoice {\hbox{$\sf\textstyle Z\kern-0.4em Z$}}
{\hbox{$\sf\textstyle Z\kern-0.4em Z$}}
{\hbox{$\sf\scriptstyle Z\kern-0.3em Z$}}
{\hbox{$\sf\scriptscriptstyle Z\kern-0.2em Z$}}}}

\newcommand{\be}{\begin{equation}}
\newcommand{\ee}{\end{equation}}

\def\fun#1#2{\lower3.6pt\vbox{\baselineskip0pt\lineskip.9pt
\ialign{$\mathsurround=0pt#1\hfil##\hfil$\crcr#2\crcr\sim\crcr}}}

\begin{document}
\thispagestyle{empty}
\noindent\hspace*{\fill}  FAU-TP3-97/7 \\

\begin{center}\begin{Large}\begin{bf}
Magnetic and Thermodynamic Stability \\ of SU(2) Yang-Mills Theory  \\
\end{bf}\end{Large}\vspace{.75cm}
 \vspace{0.5cm}
Vladimir L. Eletsky\footnote{Permanent Address:
Institute of Theoretical and Experimental Physics, Moscow 117259,
Russian Federation},
Alex C. Kalloniatis,
\\
Frieder Lenz and Michael Thies, \\
Institute for Theoretical Physics III \\
University  of Erlangen-Nuernberg \\
Staudtstrasse 7 \\
D-91058 Erlangen, Germany
\end{center}
\vspace{1cm}\baselineskip=35pt

\date{\today}
\begin{abstract} \noindent
SU(2) Yang-Mills theory at finite extension or, equivalently, at finite
temperature is probed by a homogeneous chromomagnetic field. We use a
recent modified axial gauge formulation which has the
novel feature of respecting the center symmetry in perturbation theory.
The characteristic properties of the $Z_2$-symmetric phase, an
extension-dependent mass term and antiperiodic boundary conditions, 
provide stabilization
against magnetic field formation for sufficiently small extension or
high temperature. 
In an extension of this investigation to the deconfined phase
with broken center symmetry, the combined constraints of
thermodynamic and magnetic stability are shown to yield  
many of the high temperature properties of lattice SU(2) gauge theory.

\end{abstract}

\newpage\baselineskip=18pt

\section{Introduction}

Although no direct phenomenological data is available
for hadronic physics at high temperature, Monte Carlo lattice simulations
of pure Yang-Mills theory and QCD begin to unfold a consistent picture.
In pure Yang-Mills theory, a phase transition occurs
where the Polyakov loop as order parameter signals the change from
a confined to a deconfined phase. This corresponds to
a change from the phase with unbroken center symmetry to the spontaneously
broken phase.
Moreover, there is now ample evidence
\cite{Karsch,Boyd} of nontrivial
gluonic structure above $T_c$ which persists over a significant
temperature range.
The approach to the high temperature limit of a free gluon gas
is rather slow, and even at very high temperatures standard
perturbation theory has only met with limited success.

In this paper, we seek to gain some additional insight into this
behaviour
in pure SU(2) gauge theory by exploiting those nonperturbative features
which are analytically
accessible in the modified axial gauge.
Recently, an effective action
was derived by some of us \cite{LT97} for
the situation in which one of the three space dimensions is compact
(at ``finite extension"). As a result of integrating out
the gauge field variables which describe the phase of the Polyakov loop,
this effective action has two novel features: a mass term and antiperiodic
boundary conditions in the finite spatial direction for
charged gluons.
The mass and changed boundary
conditions are themselves non-perturbative and a consequence of the
nontrivial Jacobian in this gauge.
In the same work, the
correspondence between this finite extension formulation and
the analogue system in thermodynamic  equilibrium at finite temperature
was detailed. Thus in the present work we shall use
the terms ``temperature" and ``extension" interchangeably, with
the following replacements understood,
$T \leftrightarrow 1/L$ for temperature and inverse extension, 
and, respectively,  $\varepsilon \leftrightarrow - P$  for energy density
and pressure.

The most striking feature of the modified axial gauge is the fact that
the center symmetry is realized in perturbation theory,
implying an infinite free energy of a single static quark. Concomitantly,
the correlator of Polyakov loops  produces a linear potential.
However, perturbation theory is clearly not adequate to describe the
dynamics of confinement, so that for instance the value of the string tension
is unrealistic.

To apply this formalism to
the high temperature regime
is nontrivial since a breakdown of the center symmetry has to occur.
We cannot cross the critical point by using only perturbative methods.
If one can at all think about the high temperature phase in terms of
boundary conditions and masses for otherwise weakly interacting gluons,
it is clear that
the boundary conditions of charged gluons cannot remain antiperiodic
above $T_c$, but should, at least at very large $T$, become
periodic.

A useful tool for gaining some insight beyond perturbation
theory is that of background fields. One probes a given
system with a specific external field and looks for regimes
of stability respectively instability in the
energy density, at a certain order in
the fluctuations about this external field.
For SU(2) Yang-Mills theory
a simple, if crude, choice of background field is the
Savvidy ansatz consisting of a constant colour neutral
chromomagnetic field
\cite{Sav77}. This ansatz itself has a long history at
both zero and nonzero temperatures.
Indeed, various authors have tried to exploit the
energy density or effective potential
for a Savvidy type external field
in order to observe the deconfinement phase transition,
by studying the change in the position of the
minimum of the energy density.
Results, however, have been under dispute to date due to the
varying degrees of {\it ad hoc} measures used in
these works. In the original (zero temperature) study of Savvidy,
the minimum of the energy density is driven away from the zero field
limit by a quite specific mode; the same mode however makes the
potential imaginary at the local minimum that is finally
established (to one loop), so that this minimum is in turn unstable.
At finite temperature for the standard Yang-Mills
action, the unstable mode persists:  a number of authors
\cite{MR81,Kap81} have simply
cut this mode out and obtained a nontrivial
minimum at low temperature which becomes a minimum at zero
field for high temperature. This was then interpreted as
a signal of
the deconfinement phase transition.
However, as demonstrated in  \cite{DS81,ES86}, 
even with quarks once the unstable mode is
put back in, at high temperature the instability
persists: there is  no temperature at which
a stable minimum to the one loop energy density
at zero field develops in the background field gauge.
In another approach \cite{MO97}, and one quite similar to
ours, this effective potential has been studied by 
treating the boundary conditions
of charged gluons as a variational parameter. 
For any nonperiodic choice of boundary conditions,
the corresponding shift in the lowest Matsubara frequency
is found to make the theory stable at sufficiently high temperatures.
However, with the change in boundary conditions
in the field dependent part only, 
these results are hard to interpret physically.

Our strategy will be as follows. After reviewing the
key results of \cite{LT97},
we shall study the Z$_2$ symmetric phase, that is the phase
with the center symmetry realized,  of the SU(2) Yang-Mills theory via its
response to an external magnetic field of the Savvidy type.
To this end, we shall compute
the energy density in zeta-function regularization
\cite{EORBZ94} for SU(2) gluons in the presence of the Savvidy
 background field. For later generalization we perform this calculation for
arbitrary mass and boundary conditions of the charged gluons.
 With the ``canonical" masses and
boundary conditions as derived in the modified axial gauge
theory in \cite{LT97}, the only minima
in the energy density that appear are at zero external field,
but for sufficiently high temperature these
minima are stable.
The Z$_2$ symmetric, perturbative
phase is however not and cannot be the correct
high temperature phase of the SU(2) Yang-Mills theory. If
antiperiodic boundary conditions persisted beyond the ``critical"
point, dimensional reduction would
leave a $2+1$ dimensional Maxwell theory rather than
the non-Abelian theory that is seen in lattice computations. Furthermore,
for sufficiently high temperatures where we do not expect 
significant  non-perturbative dynamics, the thermodynamic instability
of the Z$_2$ symmetric phase becomes an important issue. We therefore will
study in the final part the consequences of magnetic and thermodynamic
stability relevant for the high temperature phase. Allowing for variations
in mass and boundary conditions, it will be possible to simultaneously satisfy
both stability requirements. As displayed in a discussion of
the high temperature properties of the system,
the resulting  phenomenological description shares, qualitatively, important
characteristics with finite temperature lattice gauge theory.

\section{Review of the Modified Axial Gauge}

At finite extension or temperature, one space-time direction
is singled out; hence, an axial type of gauge is particularly convenient.
To be specific, we choose the 3-direction as the compact direction
($x_3 \in [0,L]$)
and work in Minkowski space. All our results can equally well be
reinterpreted in terms of finite temperature equilibrium
thermodynamics if desired, by going to Euclidean space and
exploiting covariance.
The other space-time directions will be denoted by $\alpha, \beta = 0,1,2$
and referred to as ``transverse". As is well known,
periodic boundary conditions imposed
on gauge fields in the 3-direction do not permit
complete elimination
of $A_3$;
a gauge invariant zero mode has to be retained. More precisely,
this 3-dimensional residual field corresponds to
the eigenphases of the field  
$W(x_{\bot})={\rm {P}} \exp (i g \int_0^{L} dx_3 A_3)$, the trace of which
can be identified with the
Polyakov loop, $w(x_{\bot})=\frac{1}{2} \mbox{Tr}\,
W(x_{\bot})$. We thus demand only the weaker, ``modified" axial gauge condition
$\partial_3 A_3 = 0$, which can be supplemented by a diagonalization in color
space, $A_3 \to a_3^3 \tau^3/2$. For simplicity, we denote
$a_3^3$ as $a_3$ in the following. The Polyakov loop is then given by
\begin{equation}
w \left(x_{\perp}\right)= \cos\left(g L a_{3}\left(x_{\perp}
\right)/2  \right) \ ,
\label{FE6}
\end{equation}
showing that in the modified axial gauge this important
variable appears as an elementary rather than composite field.

The presence of the three dimensional field $a_{3} (x_\perp)$
indicates the necessity for a further gauge fixing. Indeed, gauge ambiguities
are still present in the sector of transverse ($\alpha = 0, 1, 2$), neutral,
$x_{3}$ independent gauge fields.  Due to their Abelian nature, these
ambiguities can easily be resolved by adding a standard gauge fixing term
to the Lagrangian.
The 3-dimensional Feynman gauge
is particularly convenient for perturbative calculations,
\begin{equation}
{\cal L}_{\rm gf}  = - \frac{1}{2 L^{2}}
\left(\int_{0}^{L}dx_{3}\partial ^{\alpha}A^{3}_{\alpha}
\left(x\right)\right)^{2} .
\label{FE7}
\end{equation}
Having fully specified the gauge, we arrive at the generating functional
\begin{equation}
Z\left[J,j_{3}\right] = \int D\left[A_{\alpha}\right] d\left[a_{3}\right]
\exp \left\{
  i S_{\rm YM} \left[A_{\alpha},a_{3}\right] + i S_{\rm gf}
\left[A_{\alpha}^{3}\right]
+ i\int d^{4}x\left(A^{a}_{\alpha}  J^{a,\alpha} + a_{3}  j^{3} \right) \right\}
\label{FE5}
\end{equation}
(see also Ref. \cite{Reinh}).
Here, $S_{\rm YM}(A_{\alpha},a_3)$ is the standard Yang-Mills action with
$A_3$ replaced by
$a_3$; the integration measure for the $a_3$ functional integral is given by
\begin{equation}
d\left[a_{3}\right] = \prod_{y_{\perp}} \sin^{2}\left( gL
a_{3}(y_{\perp})/2 \right)
\Theta \left( a_3(y_{\perp}) \right) \Theta \left( 2\pi/g L
-a_3(y_{\perp}) \right)
 da_{3}\left(y_{\perp}\right)
\label{FE8}
\end{equation}
and accounts for the fact that $gLa_3/2$
appears in the parametrization of the group manifold $S_{3} $ as
the first polar angle.
As a consequence the corresponding part of the Haar measure enters
with a finite range of integration.  In the canonical formulation,
this Faddeev-Popov determinant arises as a Jacobian modifying the
kinetic energy of the Polyakov loop variables $a_{3} (x_{\perp})$
\cite{Lenz2}.
In QED the same procedure yields the standard flat measure for
$a_{3} (x_{\perp})$.

The finite range of integration for $a_3$ in the above functional integral
precludes a conventional perturbative treatment, since the quadratic
part of the
action does not yield a Gaussian integral.
Here, we follow the
approach proposed in Ref. \cite{LT97}, which amounts to integrating out
$a_3$ completely prior to any perturbative expansion.
First, we regulate the
functional integral by a transverse lattice (lattice constant $\ell$).
Subsequently, the variable $a_3$ is shifted to $a_3'=a_3-\pi/gL$. As
a consequence, the $\sin^2$ factor in the measure (\ref{FE8})
is replaced by $\cos^2$, and the integration limits become
$[-\pi/gL,\pi/gL]$. In order to preserve the standard form
of the minimal coupling of charged gluon fields to $a_3'$,
it is advantageous to redefine
these fields as follows,
\begin{equation}
A_1 \mp i A_2 \to \exp(\pm i\pi x_3/L) (A_1 \mp i A_2) \ .
\label{FE8a}
\end{equation}
This does not change the physics, but
yields new charged gluon fields which, in contradistinction to the
neutral ones, obey antiperiodic boundary conditions.

The fact that one can integrate out $a_3$ completely
rests on the following observation:
as a consequence of the finite range of integrations,
the field $a_3$ (and hence the Polyakov loop) becomes ``ultralocal";
it does not propagate in the continuum limit \cite{LT97}.
In lattice terminology,
one finds that hopping terms are suppressed by factors
$\ell/g^2L$. Since the coupling constant is expected to ``run" with $\ell$
in the standard logarithmic manner (as will indeed be confirmed below in
the present framework)
and $L$ is a macroscopic length, this factor
goes to zero in the continuum limit.

Formally, ultralocality of the Polyakov loop field $a_{3} $
means that it can be completely integrated out. 
Introducing a lattice regularization in the transverse
directions in order to define the functional integral
over this field, one arrives at finite non-Gaussian integrals
which can be explicitly carried out. No ultraviolet divergent
expressions occur in this subsector of the theory, so that
the lattice regulator can be safely taken to zero, leaving
an effective action for the remaining degrees of
freedom ($\alpha = 0,1,2$):    
\begin{equation}
S_{\rm eff} \left[A_{\alpha}\right] =
S_{\rm YM} \left[A_{\alpha}, A_{3}=0\right]
+ S_{\rm gf} \left[A_{\alpha}^{3}\right] +  \frac{M_3^2}{2} \sum_{a=1,2}
 \int d^{4}x A^{a}_{\alpha} \left(x\right)  A^{a,\alpha} \left(x\right) .
\label{FE11}
\end{equation}
In this effective action, the Polyakov loop variable has left
its signature in the
``geometrical" mass term of the charged gluons
\begin{equation}
M_3^{2}=\left(\pi^{2}/3-2\right)/L^{2}
\label{FE12}
\end{equation}
and in the change to antiperiodic boundary conditions
\begin{equation}
A_{\alpha}^{1,2} \left(x_{\perp}, L \right) = - A_{\alpha}^{1,2}
\left(x_{\perp}, 0\right) \ ,
\label{FE12a}
\end{equation}
while the neutral gluons $A_\alpha^3$ remain massless and periodic.
The antiperiodic boundary conditions are a consequence of
the mean value of the Polyakov
loop variable, the geometrical mass comes from its fluctuations.
Notice that both of these changes in the action are independent of 
the coupling constant.
In \cite{LT97}, the center symmetry
and a properly redefined order parameter for
the confinement-deconfinement transition in this
effective theory  are discussed in detail. Here we only recall
that the expectation value of the Polyakov loop 
$\langle w(x_{\perp})\rangle$ vanishes in
zeroeth order
perturbation theory, indicating that this formulation is particularly
well suited for describing the Z$_2$-symmetric phase.

Having reviewed the essential aspects of \cite{LT97}
it is useful to look at this effective theory in a wider
context. The fact that the Polyakov loops
are indicating confinement of static quarks means that
we have a natural formalism for dealing with finite temperature gauge
theories which are still in the confined phase.
As the temperature is increased, one knows from lattice
calculations that a phase transition to a deconfined phase
happens at a certain critical temperature $T_c$; correspondingly, the
Polyakov loop order parameter acquires a non-vanishing
expectation value. We cannot expect to be able to treat
this transition perturbatively. However, we can try to extrapolate
our findings into
the deconfined phase. If there exists a similarly
adequate kind of perturbation theory for the broken
phase where gluons can be treated as weakly interacting,
it should not involve antiperiodic boundary conditions,
but rather quasi-periodic ones which approach periodic boundary conditions  
in the limit $T\to    \infty$.
This is just another way of phrasing the fact
that Polyakov loops acquire an expectation value in the deconfined
phase.
Indeed, a theory of essentially free
gluons with antiperiodic boundary conditions
for very high temperatures can be ruled out
on the grounds of dimensional reduction:
in the  limit $T\to \infty$ for any choice of
boundary conditions other than periodic, charged gluons,
exhibiting a gap in their single particle spectra,
would decouple and the resulting
dimensionally reduced theory would not be three-dimensional
Yang-Mills theory but rather Maxwell theory,
in conflict with lattice results.
Another argument comes from the sign of the Casimir
pressure, which for antiperiodic boundary conditions
would signal thermodynamic instability at high temperature,
an unacceptable result.
Finally, above the critical temperature the
effective mass of charged gluons cannot increase forever linearly with
temperature, both from dimensional reduction arguments
and in order that the Stefan-Boltzmann law can be
truly restored at very high temperatures.

To summarize,
nontrivial boundary conditions, a (magnetic) gluon mass
and the realization of center symmetry at finite temperature
are the basic non-perturbative features brought about by gauge fixing
in our formalism. Now, as the temperature is increased  
a phase transition to the broken phase must occur. At this point,
evidently the boundary condition angle and the gluon mass must change.
The physics associated with this change is computationally
difficult to describe in our approach, but
nonetheless this change in the values of these parameters
will not be immediate: namely nontrivial
values should persist above the phase transition. The simplest 
way then to characterize the symmetry broken phase
is to use these parameters, though now allowing them
to take values other than those determined above. 
The requirement of stability combined with the demand that
the Stefan-Boltzmann law be approached at high temperatures can enable us
to study the behaviour of these parameters in the very high temperature regime. 
We thus have variables $\chi$ and $M$ associated with charged gluons in
the effective theory of \cite{LT97}. The first quantity describes
the angle in the boundary condition
\begin{equation}
A^{1,2}_\alpha(x_\perp,L) = e^{i\chi} A^{1,2}_\alpha(x_\perp,0)
\end{equation}
with $0 \leq \chi \leq 2\pi$, the second
the mass of these gluons.
In terms of these parameters we calculate the energy density
for an external magnetic field of the Savvidy type.
We shall then set these
parameters to their values $\chi=\pi,M=M_3$
for charged gluons, $\chi=M=0$ for neutral
gluons, which characterize the $Z_2$-symmetric phase,
and investigate the stability problem. In the next step,
we shall treat these parameters in a less restrictive fashion
in order to explore the high temperature phase.

\section{Energy Density in a Background Magnetic Field}

In the above formalism where the redundant variables 
have been eliminated,
we proceed to study the stability
properties of the effective theory of Eq.~(\ref{FE11}) by probing it with a
background magnetic field.
We choose a homogeneous external chromomagnetic field
in color 3-direction, with the potential
of the form
\begin{equation}
A^a_\mu|_{\rm bg} = - \delta^{a3} g_{\mu 1} x_2 H \ .
\label{A1}
\end{equation}
Although this looks identical to Savvidy's choice \cite{Sav77}
at first glance,
we should like to point out an important difference. Traditionally,
the background  field has been introduced before gauge fixing. 
In this context, the choice of the background field to be diagonal has
no physical meaning.
In our case however, the 3-direction
has already been specified by diagonalizing the Polyakov loop variables. 
What we investigate then is the stability of the vacuum against 
homogeneous magnetic field fluctuations
which point in the same direction in color space as the Polyakov loop.

The energy density in the presence of the
external field is the sum over the single particle
energies.
Since we cannot use the popular background gauge, we must solve
the Landau level problem in the axial gauge; some details of the calculation
can be found in the appendix.
Clearly, the neutral gluons contribute only
to the Casimir ($H$ independent) part of the energy density, which will be
included later. We thus here restrict ourselves to the charged gluon
contributions.   The expressions at this
point are ultraviolet divergent and in need of
regularization which will be done here by zeta-function methods.
In this scheme, we need to evaluate the following expression for the
vacuum energy density,
\begin{eqnarray}
\varepsilon & = &  {{H^2} \over 2}
 + {{gH}\over{2\pi L}}
\mu^{2\epsilon}
\sum\limits_{s=\pm 1}\sum\limits_{n=0}^{\infty}
\sum\limits_{k=-\infty}^{\infty}
\left[2gH(s+n+1/2) +  { { (2 \pi k+ \chi)^2} \over {L^2} } +
M^2 \right]^{1/2 - \epsilon}
\ .
\nonumber \\
\label{regularenergy}
\end{eqnarray}
The first term is just the classical energy density
of the external magnetic field.
The coefficient
of the second term comes from the density of states
of Landau levels, the parameter $\epsilon$
regularising the ultraviolet behaviour of the sum.
The arbitrary scale parameter $\mu^2$ has
been introduced to
keep the dimensions correct.
The Casimir contribution from the charged gluons will
emerge in the $H \rightarrow 0$ limit of this
expression.
Eq. (\ref{regularenergy}) can
be evaluated using standard techniques \cite{EORBZ94}
the details of which we omit here. We shall give later the full
result of the summations after renormalization
of the ultraviolet singularities, for the purpose of which
it is here sufficient to just
consider the weak field limit of the result
$gH \ll M^2$,
\begin{eqnarray}
\varepsilon &=& {{H^2} \over 2} + \frac{11 g^2 H^2}{48\pi^2}
\left[\ln \left( \frac{M^2}{\mu^2}\right) - {1 \over \epsilon}
+1+\psi\left(-\frac{1}{2}\right)
-4\sum_{n=0}^{\infty} \cos(n\chi) K_0 (nML)\right] \nonumber\\
 &+& \frac{M^4}{16\pi^2}\left[\ln\left(\frac{M^2}{\mu^2}\right)
- {1 \over \epsilon} +\psi\left(-\frac{1}{2}\right)-\frac{1}{2}
\right]-\frac{2 M^2}{\pi^2 L^2}
\sum_{m=1}^{\infty}\frac{\cos(m\chi)}{m^2}K_2(mML)
\end{eqnarray}
We recognize in this the ultraviolet divergent part (pole in
$\epsilon$). This itself has pieces
dependent on $H$ and on $M,L$.
Also, the $H$ independent part of this expression
is the unrenormalized Casimir energy for two free scalar boson fields.
We consider this in some detail first now.

\subsection{Casimir energy and pressure}
Setting $H=0$ and using that $\psi(-1/2) = 2 - \gamma - 2 \ln 2$,
we obtain straightforwardly
\begin{eqnarray}
\varepsilon(H=0)
& = & -\frac{2}{\pi^2 L^4}\left[
\left( \frac{1}{2 \epsilon} + \frac{3}{4} - \frac{1}{2} \gamma
+ \ln (\sqrt{4 \pi} \mu/M) \right) (ML/2)^4 \right.
\nonumber \\
& & \left. + (ML)^2 \sum_{n=1}^{\infty} \frac{1}{n^2} \cos (n\chi)
K_2(nML) \right] \ .
\label{m5}
\end{eqnarray}
To eliminate the
divergence for $\epsilon \to 0$, we have to properly
renormalize the energy density.
If $M\sim 1/L$ and $\chi$ is $L$-independent
the energy density can be
calculated with respect to an arbitrary reference point $\tilde L$,
as follows
\begin{equation}
\varepsilon(L)=\frac{1}{L^4} \left( \tilde{L}^4 \varepsilon(\tilde{L})
- \frac{(ML)^4}{8 \pi^2} \ln \frac{L}{\tilde{L}} \right)  \ .
\label{m6}
\end{equation}
The logarithmic term in Eq. (\ref{m6}) prevents approach to
the Stefan-Boltzmann law at small $L$ and also modifies the standard
relation
between pressure and energy density (remember that $\varepsilon$ and $-P$
are interchanged as compared to the thermal situation)
\begin{equation}
P(L)=-\frac{\partial}{\partial L} (\varepsilon L) =3 \varepsilon(L)
+\frac{(ML)^4}{8 \pi^2 L^4} \ .
\label{m7}
\end{equation}
If on the other hand we would simply ignore
the mass, but work with antiperiodic boundary conditions $\chi = \pi$,
the result would be finite
but exhibit a repulsive Casimir effect, signalling thermodynamic
instability of the corresponding finite temperature system,
\begin{equation}
P(L)=3\varepsilon(L)=\frac{7}{60} \frac{\pi^2}{L^4} \ .
\label{m8}
\end{equation}
This illustrates the above mentioned difficulties, if we would
apply the ``canonical"
values of $\chi$ and $M$ as inferred from the axial gauge in the
Z$_2$ symmetric
phase also at small $L$ (high $T$).

\subsection{Field dependent part}
With the external field non-zero, the ultraviolet
divergence contains both field dependent and $M, L$ dependent
parts. The renormalization must proceed in two steps:
first a coupling constant
renormalization after identification of the renormalization
group invariant $B = g H$ is performed.
This leads to the correct SU(2) Yang-Mills beta function to one loop order,
a confirmation
that the non-perturbative aspects of the present formulation
do not spoil asymptotic freedom.
The arbitrary scale $\mu^2$ is fixed by choosing
\begin{equation}
{{11} \over {48 \pi^2}} \ln (B/\Lambda_{\rm MS}^2) \equiv
{1 \over {2 g_{\rm R}^2(\mu)} } + {{11} \over {48 \pi^2}} \ln (B/\mu^2)
\end{equation}
where $\Lambda_{\rm MS}$ is the scale parameter for SU(2) in
the minimal subtraction scheme.
There remain now poles independent of $B$, but
dependent on $M,L$. These are eliminated by performing a
subtraction at $B = 0$.
Finally, for arbitrary field $B$
we obtain the rather lengthy renormalized expression:
\begin{eqnarray}
\varepsilon_{\rm R}(L,B) & \equiv & \varepsilon(L,B) - \varepsilon(L,0)
\nonumber \\
& = &
\frac{11 B^2}{48\pi^2} \ln\left(\frac{2B}{\Lambda_{\rm MS}^2}\right)
+ \frac{M^4}{16 \pi^2}
 \ln\left(\frac{2B}{M^2}\right)
\nonumber \\
& + &  \left[ \frac{B(M^2 + B)}{8 \pi^2}
\left( \ln\left(\frac{M^2 + B}{2|B|}\right) +
\psi(-1/2) \right) + (B \rightarrow -B) \right]
\nonumber \\
\nonumber \\
&+&
{\frac{B^2}{2\pi^2}
\left[\zeta^{\prime}\left(-1, \frac{M^2+B}{2B}\right)-\psi(-1/2)
\zeta\left(-1, \frac{M^2+B}{2B}\right)  \right]} \nonumber\\
&-&  \frac{2B}{\pi^2 L}\sum\limits_{n=0}^{\infty}[B(2n+1)+M^2]^{1/2}
\sum\limits_{m=1}^{\infty}\frac{\cos(m\chi)}{m}
K_1 \left(mL [B(2n+1)+M^2]^{1/2}\right)
\nonumber \\
&+&  \left[
\frac{B(M^2 + B)^{1/2}}{\pi^2 L}\sum\limits_{n=1}^{\infty}
\frac{\cos(n\chi)}{n} K_1 (nL(M^2 + B)^{1/2})
+ (B \rightarrow - B) \right]
\nonumber \\
\nonumber \\
&-&  \frac{M^4}{16 \pi^2}
\left( \psi(-1/2) -1/2 \right) + \frac{2M^2}{\pi^2 L^2}
\sum_{m=1}^{\infty}\frac{\cos(m\chi)}{m^2}K_2(mML)
\label{A20}
\end{eqnarray}
with the Riemann zeta-function 
$ \zeta(s,\nu)$, \cite{GR},
and $\zeta^{\prime}(s,\nu)=\partial \zeta(s,\nu)/ \partial s$.
This result is more involved than that of
previous authors mainly due to the non-vanishing gluon mass.

We consider two limits of this renormalized expression,
the weak field ($B \ll M^2$) or the strong
field case ($B \gg M^2$).
Taking the strong field limit first, we obtain
\begin{equation}
\varepsilon_{\rm R}(B \gg M^2) \approx
\frac{11 B^2}{48\pi^2}
\ln\left(\frac{B}{\Lambda_{\rm MS}^2}\right)
\label{strongfield}
\end{equation}
with subleading terms having an imaginary part now. This
is precisely the case considered by Savvidy and many others.
In the weak field case, we obtain
\begin{equation}
\varepsilon_{\rm R}(B \ll M^2) \approx
\frac{11 B^2}{48\pi^2}
\left[\ln\left(\frac{M^2}{\Lambda_{\rm MS}^2}\right)
+1+\psi(-1/2)
-4\sum_{n=1}^{\infty} \cos(n\chi) K_0 (nML)\right]
\label{weakfield}
\end{equation}
Note that due to the existence of the additional
mass scale, we no longer have the magnetic field
appearing in the logarithm. This has important consequences for the
discussion of magnetic stability.

We now input the ``theoretical" values
characteristic for the Z$_2$ symmetric phase
for the parameters $M$ and $\chi$, i.e.
$M=M_3=(\pi^2/3 -2)^{1/2}/L, \, \chi = \pi$
(note that only the charged gluon contributions
have survived after the zero-field subtraction).
We have numerically computed $\varepsilon_{\rm R}$ for
various values of the extension $L$ against $B$.
As a result, we find no sign of minima at
nonzero $B$-field values for which
there is no imaginary part. Only
at the point $B=0$ can a stable minimum occur. Comparing the numerical
to the weak field result demonstrates the
effectiveness of the lowest order $B$ contribution
in the neighborhood of the minimum. This is shown in
Fig. 1. For larger values of $L$, the energy density would start with
a negative curvature and the cusp visible in Fig. 1 be more
pronounced. At the cusp, the energy develops an imaginary part;
this will be discussed further in the following subsection.

Having established that $B=0$ represents the only
relevant minimum, it is simple to determine the critical
value of $L$ below which the sign of $B^2$ in Eq. (\ref{weakfield})
is positive for the canonical values for $\chi$ and $M$.
It is found to be
\begin{equation}
L_c  =  {{L M_3} \over {\Lambda_{\rm MS}}} \exp[ (1 + \psi(-1/2) -
 4 \sum_{n=1}^\infty (-1)^n K_0(nL M_3))/2] \
       \approx   {{3.36} \over {\Lambda_{\rm MS}}} \ .
\end{equation}
For magnetic stability,
one must have $L \leq L^c$. For these values of $L$
the system can be said to be preferring the
``empty" vacuum $(B=0)$, even though the calculation does not yet
contain enough dynamics to make statements about the
true small $L$ (high temperature) behaviour.
Given that $T_c \approx \Lambda_{\rm MS}$,  
this means that the center-symmetric phase is stable above roughly
$T_c/3$. Thus, anticipating the final section, it is plausible to assume 
that in the broken phase this 
stability will persist.

\subsection{Imaginary part of energy density}

Above, we have focussed on the sign change of the $B^2$ contribution
to the renormalized energy density, i.e., the weak field aspect.
Our calculation shares with previous
investigations the disease that above a certain critical magnetic field,
the energy density develops an imaginary part. In fact, this happens
at all values of $L$. In Fig. 1, it is reflected in the cusp in the
real part of $\varepsilon$, whose position
corresponds to the threshold magnetic field;
the curves at smaller
$L$ develop such cusps at larger $B$-values not visible in the figure.
The value of the critical magnetic field can easily be understood:
this is nothing
but the point at which the mass gap (due to the mass
and nonzero lowest Matsubara mode) is completely compensated
for by the magnetic field
(the lowest Landau level of the
$s=-1$ gluon reaches zero,
see Eq. (\ref{regularenergy})).
This kind of instability at large magnetic fields is more problematic
than the one at weak fields, since we cannot trust the one loop
calculation anymore.
Clearly, at the critical
magnetic field, a restructuring of the vacuum must take place
since the gluons in the lowest Landau level can condense.
Here, we have made no attempt to treat these particular
modes non-perturbatively and therefore cannot handle
this kind of instability.
One interesting aspect however
is worth mentioning, which is specific for our approach: as elaborated
in ref. \cite{LT97}, the Polyakov loop correlator is dominated
at large distance by the closest singularity in momentum space,
i.e. the threshold for producing two charged gluons. On the other hand,
it is related to the potential between static fundamental sources.
On the basis of the Landau orbits, we can immediately predict that the
``string tension" will go like
\begin{equation}
\kappa  =  2(c^2/L^2 - 2 B)^{1/2}
\label{stringconst}
\end{equation}
with $c^2 = 4 \pi^2/3 - 2$ coming from the
combination of the mass term and the lowest Matsubara mode.
At the cusp, the string tension vanishes; this opens
the possibility for exploring the center symmetry breaking
as a function of a parameter other than the temperature, namely
the strength of the background field. We refrain from doing
this here since it would require a better dynamical treatment of those
modes which might condense.

\section{High Temperature Regime}

We appear to have arrived at a partly satisfactory description
of the response of the Z$_2$ symmetric Yang-Mills phase to a
homogeneous magnetic field at some intermediate temperature.
As emphasized above, extension towards high temperatures is
problematic. Although increasing the temperature improves the
stability with respect to magnetic field fluctuations, the
thermodynamic instability now becomes an important issue.
Criteria of thermodynamic stability 
generally imply second order derivatives of thermodynamic quantities. 
We expect at sufficiently high temperature a gas of gluons
in a regime where the Stefan-Boltzmann law is approximately
satisfied, namely the $T^4$ dependence of energy density or
pressure is dominant. Thus our simplified criterion for stability 
will be {\it positive pressure} for thermodynamics, or
equivalently {\it negative energy density} for finite extension.  

Now, the requirements of stability against magnetic field
formation and thermodynamic stability are to some extent in
conflict. Roughly speaking, magnetic (thermodynamic)
stability demands large (small) values of
the ``infrared parameters" $M, \chi$.
It is thus of interest to discuss the consequences of the
combined constraints of magnetic and thermodynamic stability.
This study will be performed by considering  variations in
the quantities $M$ and $\chi$ in order to achieve  simultaneously
stability with respect to both thermal and magnetic field fluctuations.
To allow for variations in these parameters is clearly appropriate
for a description of the high temperature phase. At
temperatures or extensions beyond the confinement-deconfinement transition
with the center symmetry spontaneously broken, the quantities $M$ and $\chi$
must deviate from their canonical values characteristic for the Z$_2$
symmetric phase. The aim of our studies is to display the consequences
of the stability rather than to attempt a fit to lattice
results for the high temperature phase. We will not allow for
simultaneous parameter changes, but identify one of the two parameters
with its infinite temperature limit ($M=0$ or $\chi=0$).
In the first place, we shall assume massless charged
gluons with $L$-dependent angle $\chi$ (quasi-periodic gluons);
then, we shall
keep the boundary conditions of the charged gluons
periodic but allow for a more general $L$-dependence
of their mass (massive gluons).

\subsection{Stability with quasi-periodic gluons}

Here we assume $ML$ to be negligible. We first consider the requirement of
thermodynamic stability. The Casimir energy density for the massless
charged gluons with quasi-periodic boundary conditions can be deduced from
Eq. (\ref{m5}),
\begin{equation}
\varepsilon = - \frac{4}{\pi^2 L^4}
\sum_{n=1}^{\infty}\frac{\cos (n\chi)}{n^4}
 =  \frac{4\pi^2}{3L^4} B_4\left( \frac{\chi}{2\pi} \right) \ ,
\label{m13}
\end{equation}
with the Bernoulli polynomial $B_4(x)=-1/30+x^2(1-x)^2$.
Eq. (\ref{m13}) displays the Stefan-Boltzmann behaviour
for periodic boundary conditions
as well as a change of sign for antiperiodic ones. The Casimir pressure
is given by
\begin{equation}
P=-\frac{\partial (L\varepsilon)}{\partial L} = 3 \varepsilon -
\frac{8 \pi}{3 L^3} B_3\left( \frac{\chi}{2 \pi} \right)
\frac{\partial \chi}{\partial L} \ .
\label{m14}
\end{equation}
Positive pressure in thermodynamics corresponds to a negative Casimir 
energy density, Eq.~(\ref{m13}), and therefore requires
\begin{equation}
\chi \le 1.51 \ .
\label{m14a}
\end{equation}
A positive susceptibility is a necessary condition for magnetic
stability. The vanishing of the $B^2$ term in the energy density
for the case $ML=0$
can be determined by
carrying out the sum in Eq. (\ref{weakfield}) with the help
of the identity
\begin{eqnarray}
\sum_{n=1}^{\infty} \cos(n\chi) K_0(ny) &=& \frac{1}{2}\left(\gamma+
\ln \frac{y}{4\pi}\right) + \frac{\pi}{2\sqrt{y^2+\chi^2}} +
\label{identity} \\
&+ &  \frac{\pi}{2} \sum_{n=1}^{\infty} \left( \frac{1}{\sqrt{y^2+
(2\pi n-\chi)^2}} + \frac{1}{\sqrt{y^2+(2\pi n +\chi)^2 }}-\frac{1}{\pi n}
\right)  \ .
\nonumber
\end{eqnarray}
The requirement that expression
(\ref{weakfield}) vanishes then simplifies to
\begin{equation}
L \Lambda_{\rm MS} = 4 \pi \exp \left\{ \frac{1}{2}
\left(1-\frac{2\pi}{\chi}
+\psi(-1/2)+\psi\left(1+\frac{\chi}{2\pi}\right)
+ \psi\left(1-\frac{\chi}{2\pi}\right) \right) \right\} \ .
\label{m9}
\end{equation}
This relation, together with the inequality (\ref{m14a}), divides the
($\chi, L$) plane into magnetically stable and unstable
regions as seen in Fig. 2. Notice that the combined constraints from
magnetic and thermodynamic stability are quite restrictive.
Inverting Eq. (\ref{m9}) yields a lower bound
for $\chi$ as function of $L$. At small $L$ in particular, this inversion can
be performed in closed form
with the result
\begin{equation}
\chi (L\to 0) \approx  \frac{2\pi}{3(1-\gamma)+2 \ln(2\pi/(L\Lambda_{\rm MS}))} \ .
\label{m10}
\end{equation}
It is instructive to reinterpret this formula as high $T$ behaviour and
eliminate the logarithm in favor of the (one loop) running coupling
constant,
\begin{equation}
g^{-2}(T) = \frac{11}{12 \pi^2} \ln(2\pi T/\Lambda_{\rm MS}) \ .
\label{m11}
\end{equation}
Then, for large $T$ we find the magnetic stability bound
\begin{equation}
\chi(T) \geq \frac{11}{12\pi} g^2(T) \ ,
\label{m12}
\end{equation}
now in a more conventional, seemingly perturbative, guise.
In the case of quasi-periodic gluons, any angle $\chi$
vanishing faster then $g^2(T)$ (that is, faster than $1/\ln(T)$)
for $T \to \infty$ would be ruled out.

In the absence of a complete theory for $\chi(L)$, we simply identify
$\chi(L)$ with its bound obtained from the requirement of magnetic
stability (Eq. (\ref{m9}) and, at large $T$, Eq. (\ref{m12})).
If we then reinterpret
$L$ as $1/T$ and interchange $\varepsilon$ with $-P$ (and vice versa),
we can get an upper limit for pressure and energy density of a gluon gas
at high temperature, at least within the limited set of
configurations which we can describe.
This has been done in Fig. 3, where $\varepsilon$
and $P$ are normalized to the Stefan-Boltzmann values $\varepsilon=3P=
\pi^2 T^4/5$ and the contribution from neutral gluons (which we assume
to obey periodic boundary conditions) has been
added.

We first note that we cannot perform such a calculation at temperatures
below $T=0.34 \Lambda_{\rm MS}$, simply because Eq. (\ref{m9}) ceases to have a
solution (the right hand side is bounded from above).
At low temperatures in Fig. 3, we observe the onset of thermodynamic
instability where the pressure changes sign (or the Casimir effect
changes from attraction to repulsion) because $\chi$ crosses
the value 1.51, see Eq. (\ref{m14a}).
The most conspicuous feature of Fig. 3 is however the slow,
logarithmic approach to the Stefan-Boltzmann limit, reminiscent of
the lattice data \cite{EKSM82} but
quantitatively farther away from this limit
in the covered temperature region.
A rather sensitive measure for interaction effects is the quantity
$\varepsilon - 3 P$, which is of interest also in view of
its direct relation to the trace of the energy-momentum tensor and the gluon
condensate \cite{Boyd}. Although our predicted temperature dependence
is too rapid, this
difference has the right order of magnitude as compared to the lattice
data of \cite{Boyd}
as illustrated in Fig. 4 (curve labelled ``$\chi$"); we have
assumed $\Lambda_{\rm MS} \approx T^{\rm{lattice}}_c \approx 290$ MeV 
\cite{EKSM82}
for this particular comparison.
We finally note the asymptotic behaviour of the thermodynamic functions
shown in Fig. 3,
\begin{equation}
\frac{\varepsilon(T)}{\varepsilon_{\rm SB}(T)} = \frac{P(T)}{P_{\rm SB}(T)}
\approx 1 - \frac{5}{3} \left(\ln
\frac{2\pi T}{\Lambda_{\rm MS}}\right)^{-2}
\qquad (T\to \infty) \ .
\label{m15}
\end{equation}
The difference $\varepsilon - 3 P$ on the other hand behaves
like $T^4 \left( \ln(2\pi T/\Lambda_{\rm MS})
\right)^{-3}$ for large $T$.

\subsection{Stability with massive gluons}
In this section,
we shall set $\chi = 0$ for all $L$ and
study the effect of an $L$-dependent, ``magnetic'' gluon mass.
The condition for the
critical mass above which the system is magnetically stable reads
(see Eqs. (\ref{weakfield},\ref{identity}))
\begin{equation}
2 \ln \frac{L\Lambda_{\rm MS}}{4\pi} = 1 -2 \gamma +\psi(-1/2)-
\frac{2\pi}{ML} - 2\pi \sum_{n=1}^{\infty} \left( \frac{2}
{\sqrt{(ML)^2+(2\pi n)^2}}-\frac{1}{\pi n} \right) \ .
\label{m16}
\end{equation}
The (numerically obtained) boundary between stable and unstable regions
in the $(M,L)$-plane for this massive gluon case is shown in Fig. 5.
Stability is only possible
for values of $M$ larger than a certain limiting value,
\begin{equation}
M_0 =\Lambda_{\rm MS} \exp \left(-\frac{1}{2}(1+\psi(-1/2)) \right)
\approx 0.60\Lambda_{\rm MS} \ .
\label{m17}
\end{equation}
The small $L$ behaviour of $M$ can again be derived analytically,
\begin{equation}
M(L \to 0)  \approx \frac{2\pi}{L(3(1-\gamma)+2 \ln (2\pi/L \Lambda_{\rm MS}))}
\ .
\label{m18}
\end{equation}
Inserting as above the running coupling constant and interchanging $L$ and
$1/T$, this corresponds to the lower bound
\begin{equation}
M(T) \ge \frac{11}{12 \pi} T g^2(T)   \qquad (T \to \infty)
\label{m19}
\end{equation}
at high temperatures.
It is remarkable that this coincides with both theoretical
expectations and lattice results for the temperature dependence of the 
gluon magnetic mass; thus
for instance, a recent determination in the Landau gauge in a wide
temperature range \cite{HKR95} has been fitted with the formula
\begin{equation}
M_{\rm m}(T)=(0.46 \pm 0.01) T g^2(T) \ ,
\label{m20}
\end{equation}
surprisingly close to (but larger than) our lower bound from magnetic
stability, Eq. (\ref{m19}). We emphasize that we have obtained Eq. (\ref{m19})
using perturbative methods, although the magnetic gluon mass is well known to
be a genuinely non-perturbative quantity. This is obviously a result from
not calculating $M$ directly, but inferring it indirectly from magnetic
stability considerations.

We can again estimate thermodynamic functions
by inserting the critical mass into the expressions for the Casimir effect
above, Eq. (\ref{m5}).
The important feature here is that the gluon mass rises more slowly
than linearly
with temperature. This is highly welcome, since it guarantees that we
will asymptotically recover Stefan-Boltzmann behaviour of an ideal gluon gas.
Since on the other hand, magnetic stability does not
permit a mass growing more slowly than with $T/\ln(T)$, these
restrictions leave very little room for a different behaviour.
In contrast to the massless case with $T$-dependent
boundary conditions (quasi-periodic gluons), we now
need one subtraction to get a finite result,
see Eq. (\ref{m5}).
This reduces somewhat the predictive power of the approach based
on massive gluons.
Nevertheless,
encouraged by the reasonable $T$-dependence of the gluon mass, we
proceed, adjusting the energy
density at one reference length $\tilde{L}$. After this subtraction,
we have to evaluate the Casimir energy density for charged gluons
\begin{equation}
\varepsilon= M^4\left(\frac{\tilde{\varepsilon}}{\tilde{M}^4} +
\frac{2}{16\pi^2} \ln \frac{M}{\tilde{M}} - \frac{2}{\pi^2}
\sum_{n=1}^{\infty} \frac{1}{n^2} \left( \frac{K_2(nML)}{(ML)^2}
-\frac{K_2(n\tilde{M}\tilde{L})}{(\tilde{M}\tilde{L})^2} \right)
\right) \ ,
\label{m21}
\end{equation}
where $\tilde{M}=M(\tilde{L}), \tilde{\varepsilon}=\varepsilon(\tilde{L})$.
As a sample calculation,
we set $\tilde{\varepsilon}$ equal to a certain fraction of the
Stefan-Boltzmann value (0.8) at the highest temperature; this
value has been selected since it is consistent with the
reduction of $30\%$ at $2T_c$ reported in the SU(2) lattice calculation
of \cite{EKR95}.
The pressure is evaluated by numerical differentiation, see Eq. (\ref{m7}).
The result is shown in Fig. 6 where
a massless neutral gluon contribution has been added, since our formalism
gives no hint to modifications in the neutral sector.
While different in detail,
this particular calculation leads qualitatively to a picture similar to
the one with quasi-periodic gluons shown in Fig. 3.
Once again we observe the slow (logarithmic) approach
to the Stefan-Boltzmann limit as $T \to \infty$.
We have also calculated the interaction measure $\varepsilon - 3 P$
in the region where lattice data \cite{Boyd} are available
and included it
into Fig. 4 (curve labelled ``$M$") to test the sensitivity of
these calculations to details.
The same renormalization condition as in Fig. 6 has been used
here.

\subsection{Electric screening mass}
It is tempting to use the function $\chi(T)$ derived from the condition
of magnetic stability also to improve other perturbative predictions.
One key quantity in hot gauge theories is the Debye screening mass,
related to the zero momentum limit of the vacuum polarization tensor
$\Pi_{00}$. Standard (finite extension) perturbation theory in the axial gauge
gives for $\Pi_{33}$ the
lowest order expression
\begin{equation}
m_{\rm el}^2 = - \frac{2 g^2}{\pi L^2} \sum_n \left( (2\pi n+ \chi)^2
+(ML)^2 \right)^{1/2} \ ,
\label{m15a}
\end{equation}
where the gluon loop and tadpole are included and dimensional regularization
has been used to perform the 3-dimensional integrations.
In the limit $M \to 0$ and evaluating the divergent sum with zeta function
methods, we find
\begin{equation}
m_{\rm el}^2 = \frac{4 g^2}{L^2} B_2\left( \frac{\chi}{2\pi} \right)
\label{m15b}
\end{equation}
($B_2(x)=1/6-x(1-x)$). Incidentally, this result can be checked
easily by taking the second derivative of $\varepsilon$, Eq. (\ref{m13}),
with respect to $\chi$.
With the special case of periodic boundary conditions
and replacing $L$ by $1/T$,
we recover the well known result $m_{\rm el}^2=2g^2T^2/3$, where $g^2$
should be replaced by the running coupling constant. On the other hand,
insertion of the critical value for $\chi(T)$,
the $\chi$-dependent factor in Eq. (\ref{m15b}) changes the temperature
dependence of $m_{\rm el}$ significantly. Over
a wide range of temperatures, we observe a compensation
between the decrease due to the running coupling constant and the increase
due to the restoration of periodic
boundary conditions; see Fig. 7 (curve labelled ``$\chi$").
The resulting
Debye mass has a linear $T$-dependence, $m_{\rm el} \approx  0.84 T$,
to very good accuracy from $T=4\Lambda_{\rm MS}$
up to $T$ much larger than 16$\Lambda_{\rm MS}$ shown in the plot. This
temperature dependence is consistent with the flat behaviour
observed in the lattice calculation \cite{HKR95},
although the prefactor there
was found to be significantly larger than ours, $m_{\rm el} \approx 2.5 T$.
Very recently, a new lattice calculation \cite{HKR97} reported
somewhat smaller values
of $m_{\rm el}$ in this temperature region. However, for extremely high
temperatures (up to $10^4 T_c$), the logarithmic running was observed,
although with a coefficient significantly larger than the naive
perturbative expectation.

In the other case where the magnetic gluon mass $M$ serves to stabilize
the system, we
can go through similar considerations.
Using zeta-function regularization,
$m_{\rm el}^2$ in the massive gluon case can be
treated along similar lines as the Casimir energy density. A subtraction
of $m_{\rm el}^2/M^2$ at a certain $\tilde{L}$ is necessary to
get a finite answer. We find (in analogy to Eq. (\ref{m21}))
\begin{equation}
m_{\rm el}^2 = M^2 \left( \frac{\tilde{m}_{\rm el}^2}{\tilde{M}^2}
+ \frac{g^2}{\pi^2} \ln \frac{M}{\tilde{M}} +
\frac{4g^2}{M^2} \sum_{n=1}^{\infty}\frac{1}{n} \left(
\frac{K_1(nML)}{ML} - \frac{K_1(n\tilde{M}\tilde{L})}{\tilde{M}\tilde{L}}
\right) \right)                            \ .
\label{m22}
\end{equation}
If we now
adjust $\tilde{m}_{\rm el}$ to a reasonable value, we obtain
a behaviour similar to that for quasi-periodic gluons, namely a
striking plateau in
$m_{\rm el}/T$ up to
high $T$ (Fig. 7, curve labelled ``$M$"). On the other hand,
it does not seem possible to
remove the discrepancy
in the absolute value of a factor of 2 with the lattice
data \cite{HKR95}
by simply adjusting our electric mass to a higher value at one temperature;
if we were to do this, the curve for $m_{\rm el}$ would approach the
asymptotic limit from above, and the flat
behaviour of $m_{\rm el}/T$ would be destroyed. Thus we cannot
reconcile absolute value and $T$-dependence
of the Debye screening mass better than to the shown
qualitative level.

\section{Conclusions}

In summary, we have probed SU(2) Yang-Mills theory at finite extension
by means of a homogeneous magnetic field. Technically,
our study differs in two important
respects from previous works devoted to this topic:
On the technical side, we have worked in a completely gauge
fixed framework from the outset, the modified axial gauge.
The gauge fixed theory exhibits certain non-perturbative              
ingredients like a mass and boundary condition angle 
for charged gluons in the Z$_2$-symmetric (confined) phase. 
>From the physics point of view, our investigations focussed 
then on the interplay between magnetic and thermodynamic stability. 
The calculation then proceeded in a straightforward way, by evaluation 
of the energy density to one loop order
(we prefer the Casimir ground state to a system in
thermodynamical equilibrium, but the two are equivalent). 
A tachyonic instability at low temperature against 
large fields indeed appeared in our calculation, 
But because this is in a regime where the
one-loop calculation cannot anyway be trusted, this was
not relevant for the analysis we subsequently carried out.  
Indeed, unlike in standard background gauge computations \cite{DS81,ES86},
in our gauge-fixed formulation of the Z$_2$ symmetric phase with
antiperiodic boundary conditions and the mass $\sim 1/L$ for charged gluons,
we obtained stability against weak field magnetic 
fluctuations below a certain extension or, equivalently, 
above a certain temperature. 
This temperature was found to be approximately one-third the
confinement-deconfinement transition temperature $T_c$ as determined by
lattice gauge calculations. 

The fact that our calculation achieved stability already 
before the actual transition to the $Z_2$-broken phase
gave us some confidence in extrapolating our formalism
into this broken phase. To use our formulation there  
we made one assumption: that above the phase transition
the fundamental nonperturbative quantities remain
the mass and boundary condition angle of the charged gluons.
In the course of the phase transition, 
their values should, naturally, change and of course we cannot predict
how they should change. However
a complete change to periodic boundary conditions and massless
gluons at the critical point is not possible because it would lead  
to magnetic instability.   
On the other hand, at sufficiently high temperatures
our one loop computation, with these minimal nonperturbative
modifications, should become more reliable. 
We have determined the value of the mass, or the boundary condition
angle, which satisfies in a minimal way the condition of magnetic stability.
These bounds were in turn reinserted into various thermodynamic  
quantities. We obtained features which are stongly reminiscent of 
results of lattice calculations, but which
have thus far resisted any kind of perturbative understanding. 
Specifically, we found that the slow approach of pressure and energy density
to the Stefan-Boltzmann limit, seen in
lattice results \cite{EKSM82}, within our treatment seems quite unavoidable.
Even more surprising is our success in reproducing the linear $T-$
rather than $gT-$ dependence for the electric screening mass 
seen by \cite{HKR95}. 
We finally remark that all these results do not depend
sensitively on the details of the calculation, as is best
seen by comparing the massive gluon calculations with 
those based on quasi-periodic boundary conditions.

As mentioned at the outset, the specific choice of the
Savvidy magnetic field in the gauge-fixed formulation we
have pursued in this work has quite a different meaning  
to that in background field calculations. Because the
choice of gauge has fixed both the choice of the diagonal
color direction and the orientation of the $z$-axis in space,
each possible choice of the colour and spatial orientation
of the external field represents physically different cases.
A natural extension of this work then is to perform similar
calculations for such different orientations of the chromomagnetic
field, both in internal and in Lorentz space. 
This program is nothing less than the mapping out of 
the physical directions of the effective potential.  
Similarly, it would be interesting to study, in this gauge-fixed formalism, 
radically different external fields, such as self-dual fields \cite{Leu80},
which would include both magnetic and electric fields, or
stochastic \cite{Sim95} fields. Work in these direction is in progress.

\section*{Acknowledgements}
We would like to thank
H. Griesshammer, S. Levit, S. Nedelko, M. Shifman, Y. Simonov and
U.-J. Wiese
for fruitful discussions and suggestions.
This work was supported by the grants 06ER747 and 06ER809 from the BMBF. 
V.L.E. was supported in part by the
INTAS grant 93-0283, the CRDF grant RP2-132, the RFFR grant 97-02-16131
and by the SNF grant 7SUPJ048716.

\newpage

\renewcommand{\theequation}{A.\arabic{equation}}

\section*{Appendix: Landau Levels in the Axial Gauge}

The relevant part of the effective Lagrangian is
\begin{eqnarray}
{\cal L} & = & -\frac{1}{2}\left( d_{\alpha}\Phi_{\beta}-d_{\beta}\Phi_{\alpha}
\right)^{\dagger} \left(d^{\alpha}\Phi^{\beta}-d^{\beta}\Phi^{\alpha}\right)
\nonumber \\
& & -\partial_3 \Phi_{\alpha}^{\dagger} \partial^{3} \Phi^{\alpha}
+ M^2 \Phi_{\alpha}^{\dagger} \Phi^{\alpha} + i g f^{\alpha \beta}
\Phi_{\alpha}^{\dagger} \Phi_{\beta}       \ ,
\label{l1}
\end{eqnarray}
with the
definitions ($A^{\alpha}$ denotes the color 3 component, i.e. the
background potential)
\begin{eqnarray}
\Phi_{\alpha} & = & \frac{1}{\sqrt{2}} \left( A_{\alpha}^1-i A_{\alpha}^2\right)
\ ,
\nonumber \\
d^{\alpha} & = & \partial^{\alpha} + i g A^{\alpha}
\ ,
\nonumber \\
igf^{\alpha \beta} & = &
\left[ d^{\alpha}, d^{\beta} \right]  =  ig \left(
\partial^{\alpha} A^{\beta}-\partial^{\beta} A^{\alpha} \right)
\ .
\label{l2}
\end{eqnarray}
Greek indices run from 0 to 2.
The Landau levels are most simply obtained from the classical
Euler-Lagrange equations,
\begin{equation}
\left( d_{\beta}d^{\beta} + \partial_3 \partial^3 + M^2 \right) \Phi^{\alpha}
- d^{\alpha} d_{\beta} \Phi^{\beta} + 2 \left[ d^{\alpha}, d_{\beta} \right]
\Phi^{\beta} = 0          \ .
\label{l3}
\end{equation}
Applying $d_{\alpha}$ to Eq. (\ref{l3}) and specializing to a constant
magnetic background field, we can derive the constraint
\begin{equation}
d_{\alpha} \Phi^{\alpha} = 0
\label{l5}
\end{equation}
which
simplifies the Euler-Lagrange equations to
\begin{equation}
\left( d_{\beta}d^{\beta} + \partial_3 \partial^3 + M^2 \right) \Phi^{\alpha}
+ 2 \left[ d^{\alpha}, d_{\beta} \right]
\Phi^{\beta} = 0       \ .
\label{l6}
\end{equation}
Interestingly, Eq. (\ref{l5}) which is usually
imposed as background gauge condition
can here be shown to be a consequence of the equations of motion. This is
the reason why we recover eventually the standard results, although the
background field in the gauge fixed
framework has a somewhat different meaning. Let us describe
the
color neutral, constant magnetic field in 3-direction by
the potential
\begin{equation}
A^1=-x^2 B \ , \qquad A^0=A^2=0
\label{l7}
\end{equation}
and assume $B>0$ from now on. Then, the ansatz
\begin{equation}
\Phi^{\alpha}(x) =
e^{-i \left( \omega x^0-p x^1- k x^3 \right)}\varphi^{\alpha}(x^2)
\label{l9}
\end{equation}
converts Eq. (\ref{l6}) into
\begin{equation}
\left( \begin{array}{ccc} -\omega^2+E_k^2+h_{\rm h.o.} & 0 & 0 \\
0 & -\omega^2+ E_k^2 + h_{\rm h.o.} & 2igB \\
0 & -2igB & -\omega^2 + E_k^2 + h_{\rm h.o.} \end{array} \right)
\left( \begin{array}{c} \varphi^0 \\ \varphi^1
\\ \varphi^2 \end{array} \right)
= 0
\label{l10}
\end{equation}
with
\begin{equation}
E_k = \sqrt{k^2+M^2}
\label{l11}
\end{equation}
and the harmonic oscillator Hamiltonian characteristic for Landau levels,
\begin{equation}
h_{\rm h.o.} = - \partial_{\xi}^2 + g^2 B^2 \xi^2 \ ,
\qquad \xi = x^2+\frac{p}{gB}    \ .
\label{l12}
\end{equation}
The constraint now assumes the form
\begin{equation}
-i \omega \varphi^0 + igB\xi \varphi^1 + \partial_{\xi}
\varphi^2 = 0                  \ .
\label{l13}
\end{equation}
Introducing normalized eigenfunctions of $h_{\rm h.o.}$ via
\begin{equation}
h_{\rm h.o.} R_n(\xi) = gB\left(2n+1 \right)R_n(\xi)
\equiv \Omega_n R_n(\xi)  \qquad
n=0,1,2,...
\label{l14}
\end{equation}
and using the standard algebraic treatment of the harmonic oscillator
to deal with the constraint, it is then easy to identify the following
solutions of Eqs. (\ref{l10}) and (\ref{l13}),
\begin{displaymath}
\varphi(\xi) = {\cal N} \left( \begin{array}{c} 0 \\ 1 \\ i \end{array}
\right) R_0(\xi) \ , \qquad \omega^2 = E_k^2 -gB
\nonumber
\end{displaymath}
\begin{equation}
\varphi(\xi) = {\cal N} \left( \begin{array}{c}
\sqrt{2gB} \sqrt{n+1} R_n(\xi)/\omega \\ R_{n+1}(\xi) \\
i R_{n+1}(\xi) \end{array} \right) \ , \quad \omega^2=E_k^2+gB(2n+1)
\ , \quad n=0,1,2,...
\label{l30}
\end{equation}
\begin{displaymath}
\varphi(\xi) = {\cal N} \left( \begin{array}{c}
\sqrt{2gB} \sqrt{n} R_n(\xi)/\omega \\ R_{n-1}(\xi) \\
-i R_{n-1}(\xi) \end{array} \right) \ , \quad \omega^2=E_k^2+gB(2n+1)
\ , \quad n=1,2,...
\end{displaymath}
The wavefunctions are given for the sake of completeness only.
The eigenvalues $\omega$ are infinitely degenerate with respect to $p$;
they enter into Eq. (\ref{regularenergy}) of section 3, together with
the appropriate boundary condition.

\newpage
\section*{Figure Captions}
\vskip 1.0cm

\begin{itemize}

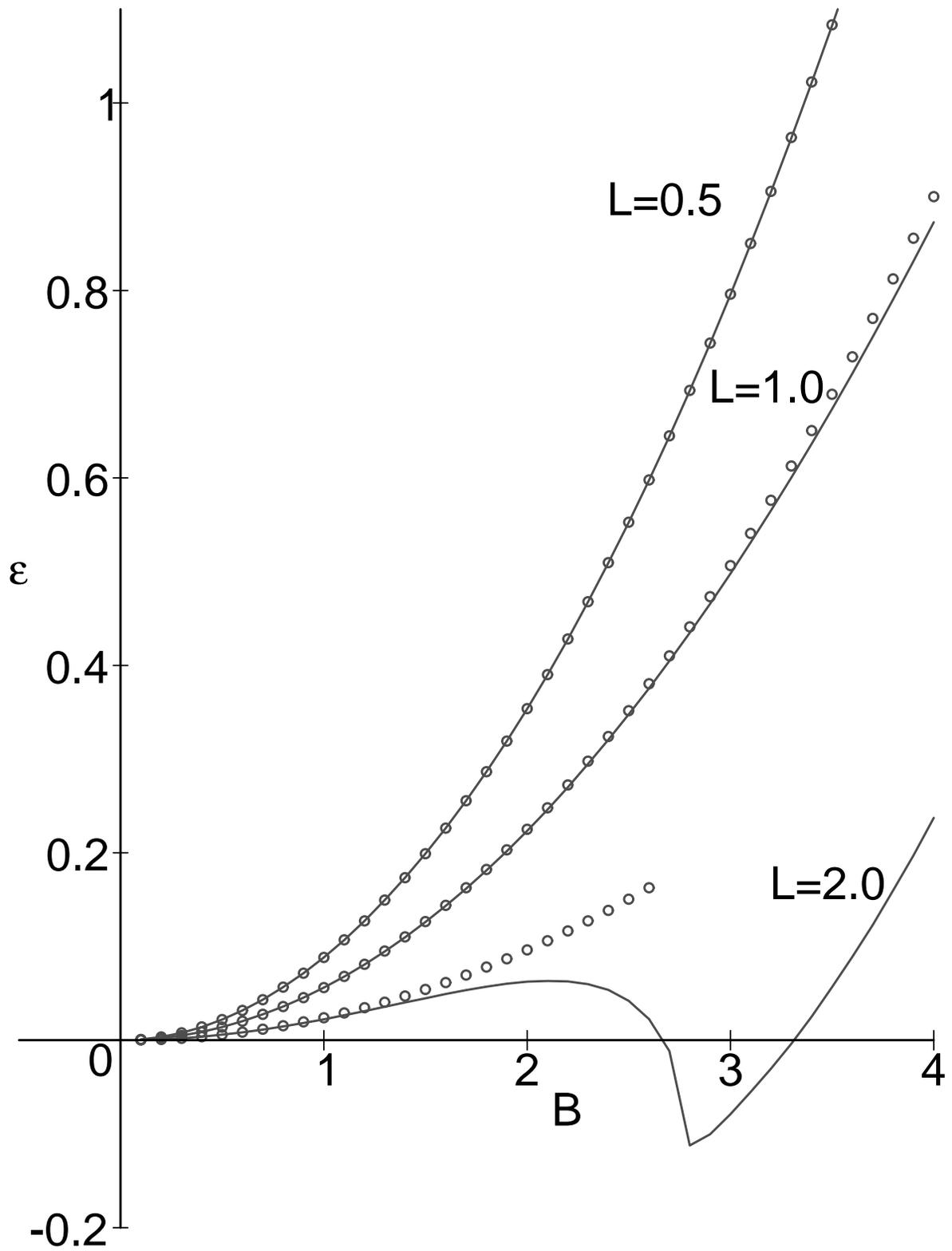
\item[{\bf Fig. 1}\ ]
Energy density of SU(2) Yang-Mills vacuum vs. magnetic field, for
3 different extension parameters $L$, in units where $\Lambda_{\rm MS}=1$.
Full curves: real part of renormalized energy density, Eq. (\ref{A20}).
Circles: quadratic approximation, Eq. (\ref{weakfield}).

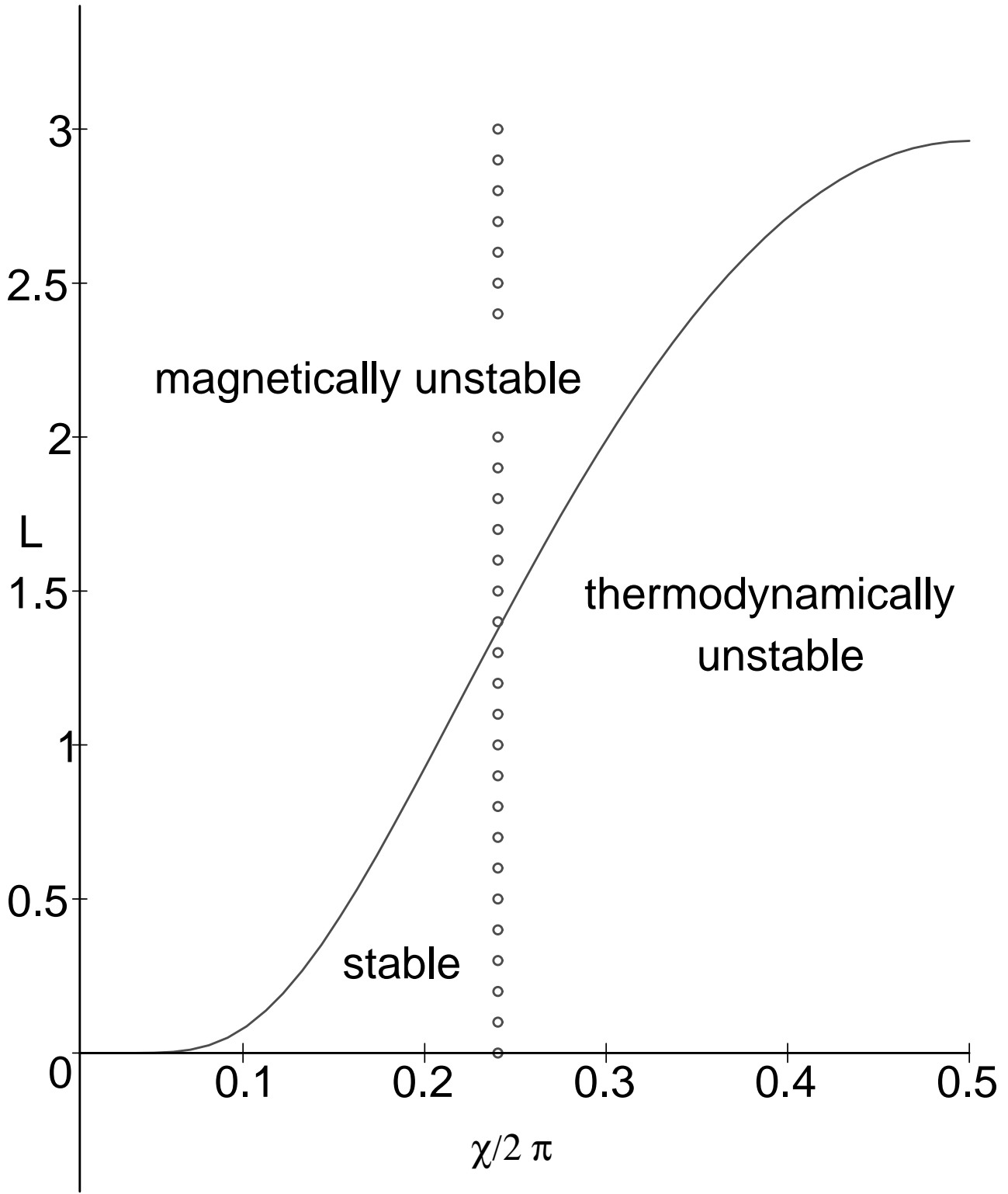
\item[{\bf Fig. 2}\ ]
Regions of stability and instability in the $(L,\chi)$ plane,
where $\chi$ is the angle for quasi-periodic boundary conditions.
To the right of
the circles, thermodynamic instability; above the solid line, magnetic
instability.

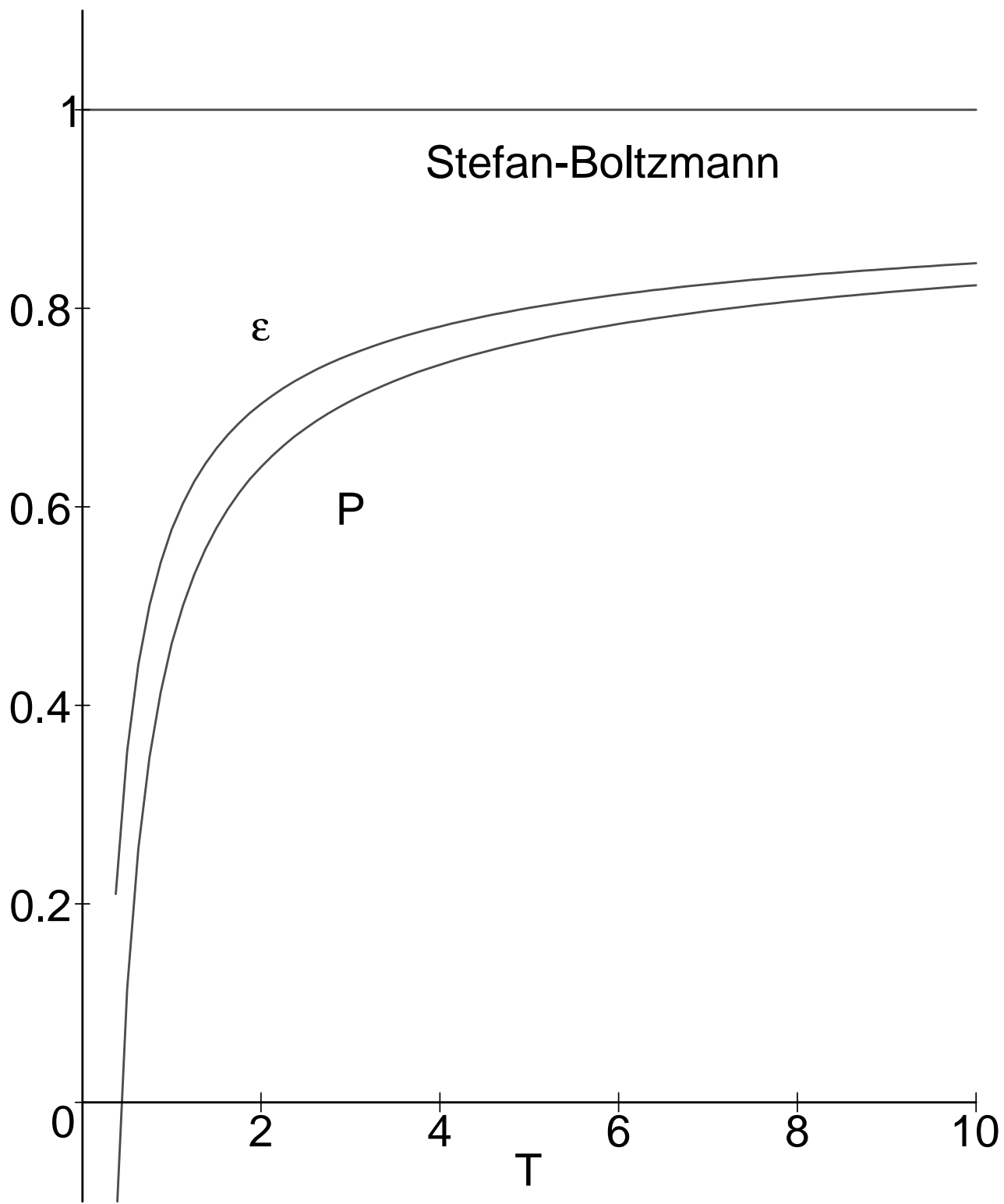
\item[{\bf Fig. 3}\ ]
Energy density and pressure, normalized to Stefan-Boltzmann values,
vs. temperature in units of $\Lambda_{\rm MS}$. The curves correspond
to the minimal model with  quasi-periodic (massless) gluons, and should
be regarded as upper bounds at high $T$.

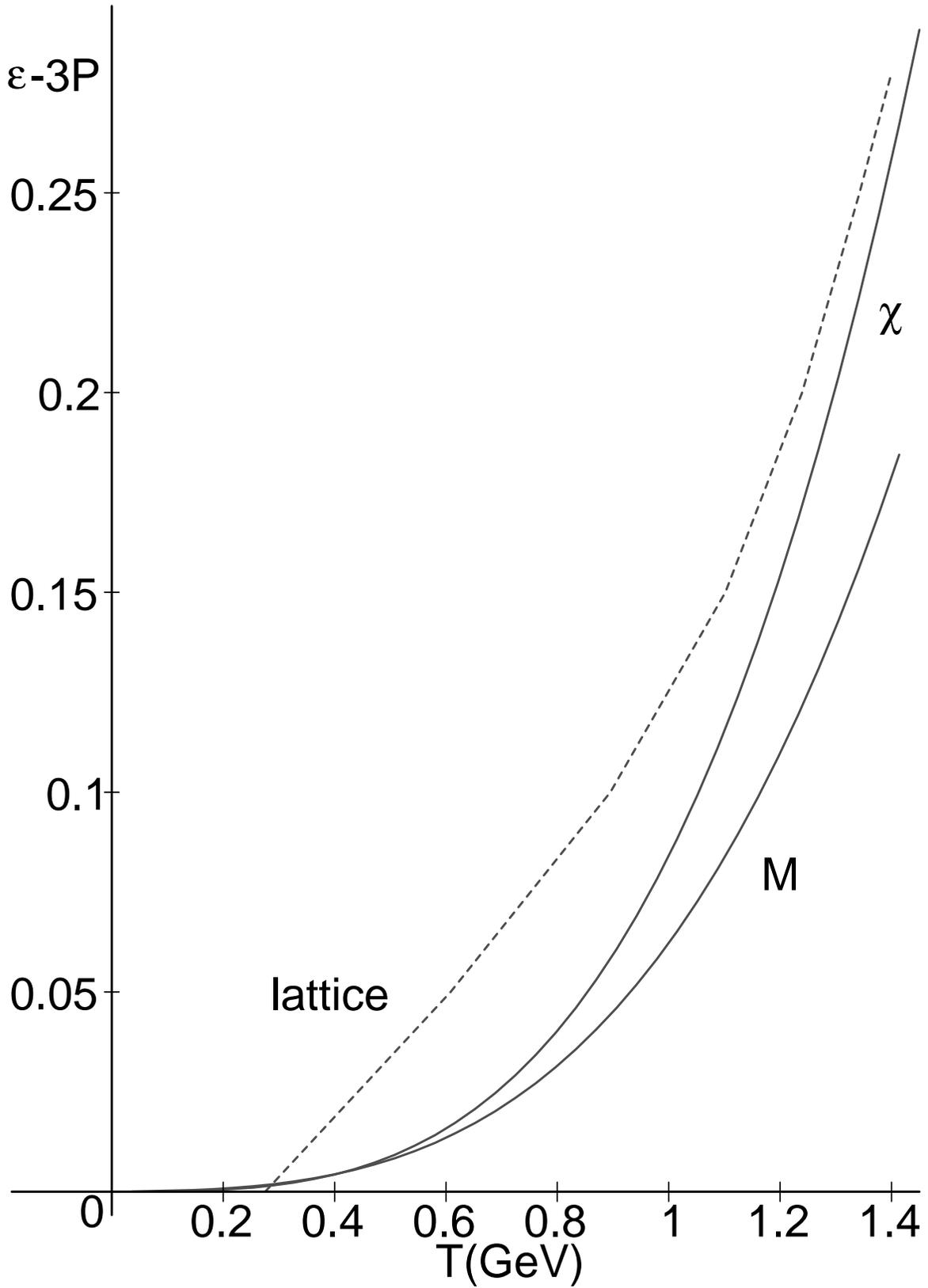
\item[{\bf Fig. 4}\ ]
Interaction measure $\varepsilon - 3 P$ vs. temperature for quasi-periodic
($\chi$) and massive ($M$) gluons, compared with lattice calculation of Ref.
\cite{Boyd}. Units have been converted to GeV.

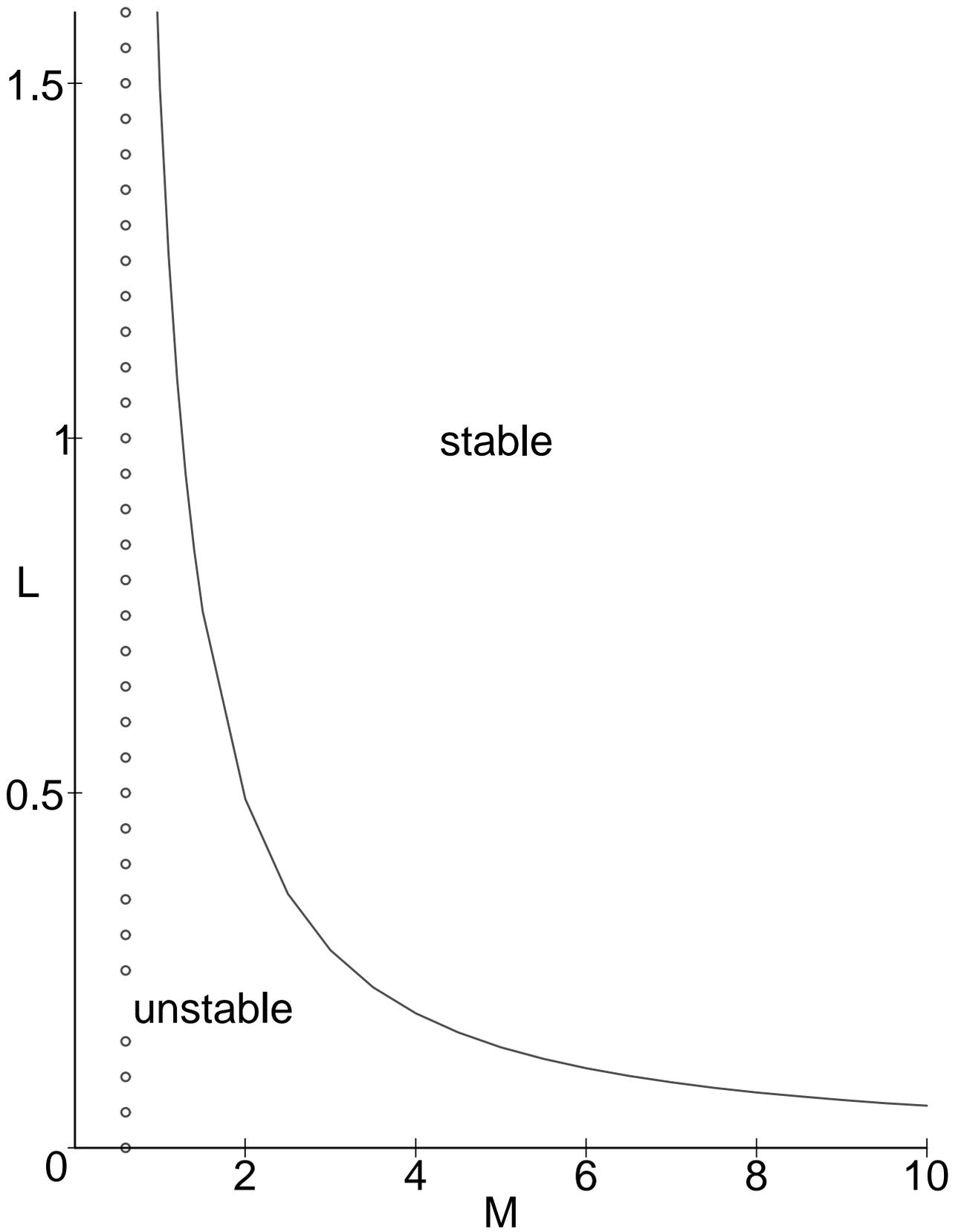
\item[{\bf Fig. 5}\ ]
Regions of magnetic stability and instability in the $(L,M)$ plane,
where $M$ is the gluon mass for massive (periodic) gluons.
Solid curve: solution of Eq. (\ref{m16}). To the left of the circles,
the system is unstable at all $L$. Units: $\Lambda_{\rm MS}=1$.

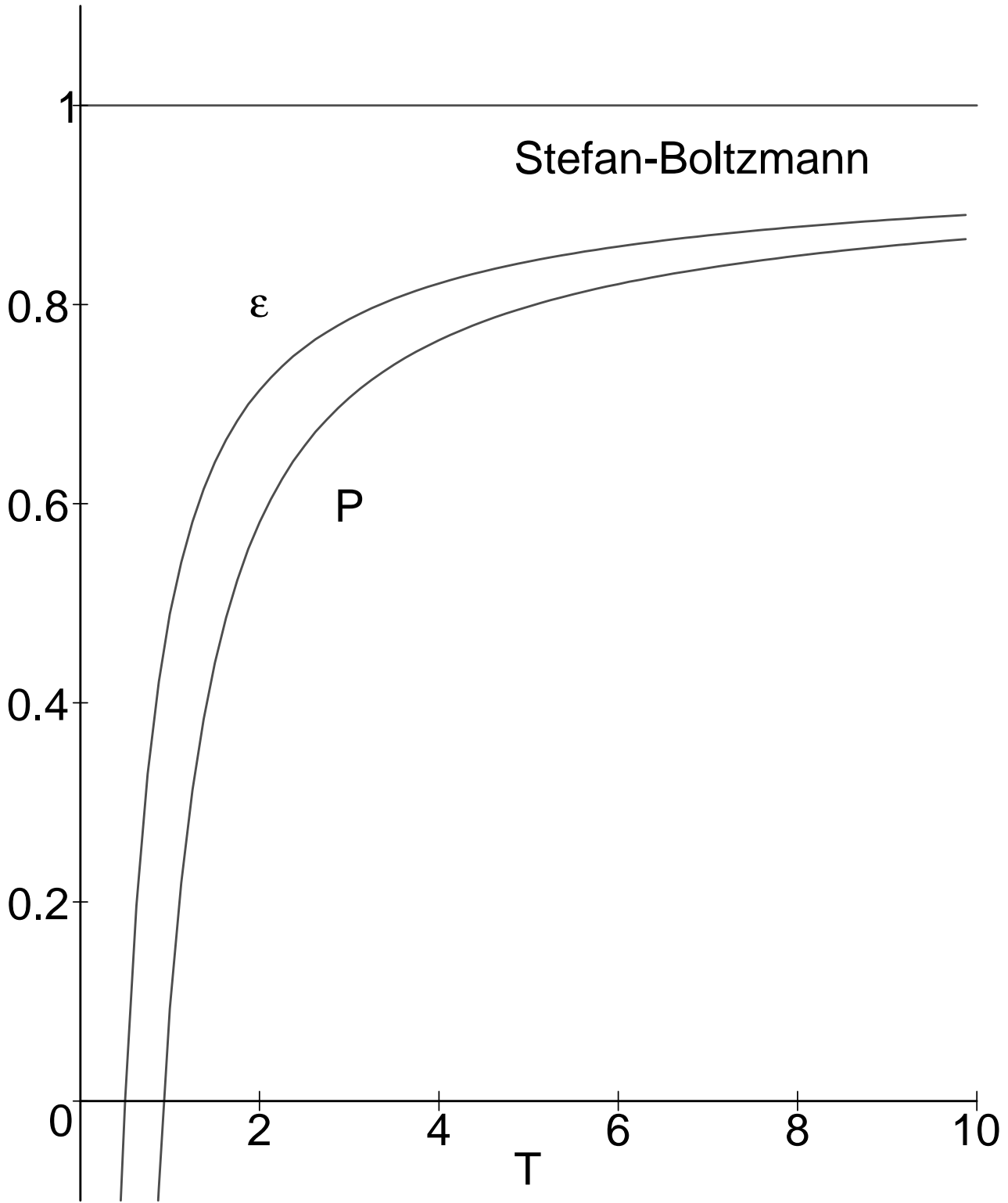
\item[{\bf Fig. 6}\ ]
Like Fig. 3, but for the minimal model with massive (periodic) gluons.

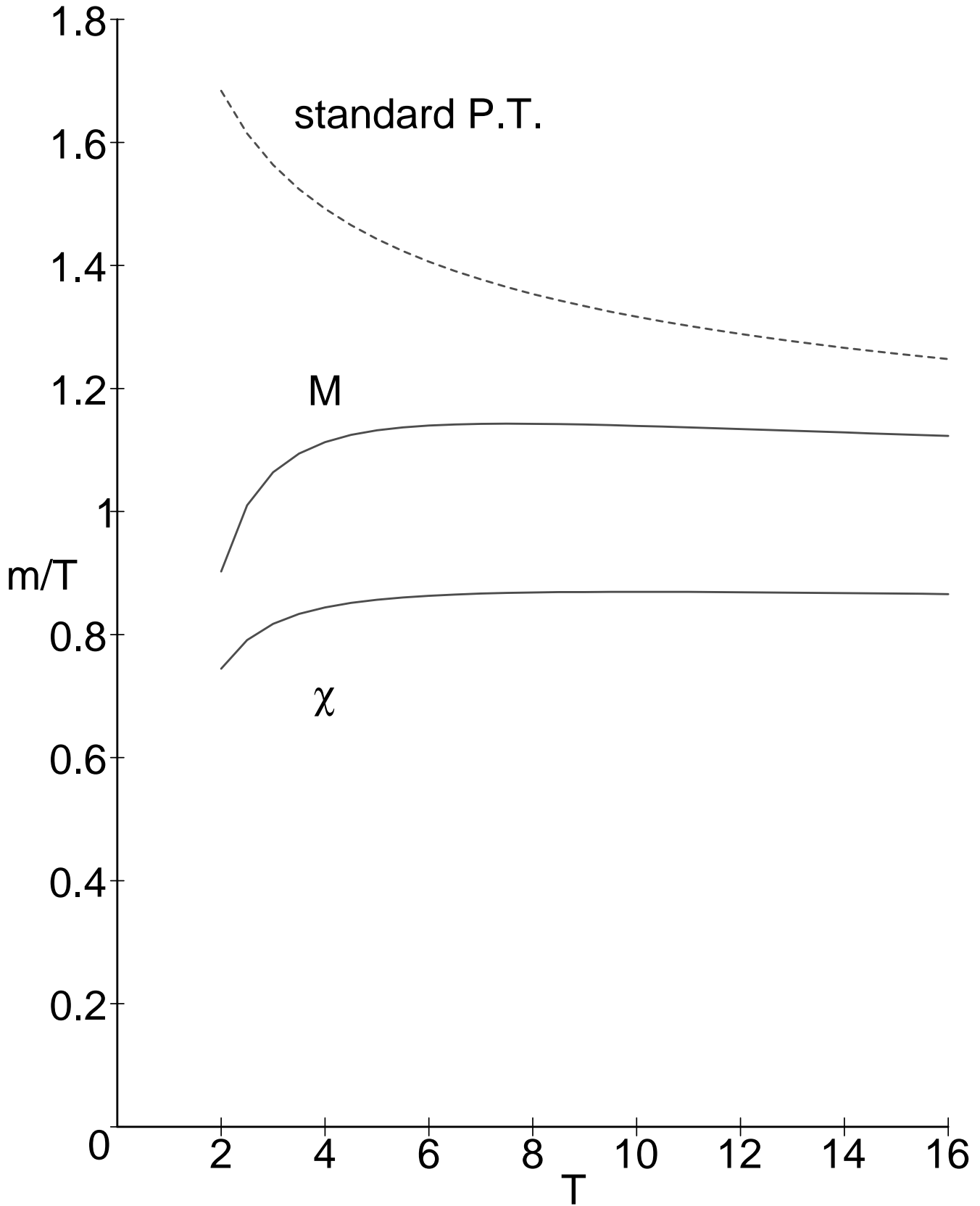
\item[{\bf Fig. 7}\ ]
Electric screening mass of gluons as function of temperature, in units of
$\Lambda_{\rm MS}$. Dashed curve: standard perturbative result with
one loop running coupling constant, $M$: massive gluons, $\chi$:
quasi-periodic gluons.

\end{itemize}

\end{document}